
\input phyzzx

\def\Z {Z}


\tolerance=10000

\def\Pp{P_{\ne}}
\def\Pm{P_=}
\def\Q{Q_+}
\def\bQ{{\bar Q}_+}
\def\bQp{{\bar Q}_{\ne}}
\def\dminus{\partial_=}
\def\dplus{\partial_{\ne}}

\def\J{\bar J}
\def\B{\bar B}
\def\I{\bar I}
\def\K{\bar K}

\def\dalemb#1#2{{\vbox{\hrule height .#2pt
        \hbox{\vrule width.#2pt height#1pt \kern#1pt
                \vrule width.#2pt}
        \hrule height.#2pt}}}

 \def\unit{\hbox to 3.3pt{\hskip1.3pt \vrule height 7pt width .4pt \hskip.7pt
\vrule height 7.85pt width .4pt \kern-2.4pt
\hrulefill \kern-3pt
\raise 4pt\hbox{\char'40}}}

\REF\Wito{E. Witten, Comm. Math. Phys. {\bf B117} (1988) 353}
\REF\Witt{E. Witten, Comm. Math. Phys. {\bf B118} (1988) 411}
\REF\Labo{J.M.F. Labastida and P.M. Llatas, Phys. Lett. {\bf B271}
(1991) 101}
\REF\Labt{J.M.F. Labastida and P.M. Llatas, Nucl. Phys. {\bf
B379} (1992) 220}
\REF\Ito{K. Ito, Phys. Lett. {\bf B250} (1990) 91}
\REF\Bir{D. Birmingham, M. Blau, M. Rakowski
and G. Thompson, Phys. Rep. {\bf 209} (1991) 129}
\REF\cmrlh{S. Cordes, G. Moore and S. Ramgoolam, "Lectures on 2D
Equivariant Cohomology and Topological Field Theories", YCTP-P11-94,
hep-th/9412210}
\REF\Hulo{C.M. Hull, G. Papadopoulos and P.K. Townsend, Phys.
Lett. {\bf B316} (1993) 291}   \REF\With{E. Witten, Nucl. Phys. {\bf B340}
(1989) 281}

\line{\hfil DAMTP R/94/55}
\line{\hfil hep-th/9412004}
\line{\hfil February 1995}

\title{\bf Topological Massive Sigma Models}
\centerline{N.D. Lambert\foot{nl10000@amtp.cam.ac.uk}}

\address{D.A.M.T.P., Silver Street\break
         University of Cambridge\break
         Cambridge, CB3 9EW\break
									England}

\vfil

\abstract

In this paper we construct topological sigma models which include a potential
and are related to twisted massive supersymmetric sigma models. Contrary to a
previous construction these models have no central charge and do not require
the
manifold to admit a Killing vector. We use the topological massive sigma model
constructed here to simplify the calculation of the observables. Lastly it is
noted that this model can be viewed as interpolating between
topological massless sigma models and topological Landau-Ginzburg models.

\endpage

%
%
\message{S-Tables Macro v1.0, ACS, TAMU (RANHELP@VENUS.TAMU.EDU)}
%
%
\newhelp\stablestylehelp{You must choose a style between 0 and 3.}%
\newhelp\stablelinehelp{You should not use special hrules when stretching
a table.}%
\newhelp\stablesmultiplehelp{You have tried to place an S-Table inside another
S-Table.  I would recommend not going on.}%
%
%
\newdimen\stablesthinline
\stablesthinline=0.4pt
\newdimen\stablesthickline
\stablesthickline=1pt
%
%
\newif\ifstablesborderthin
\stablesborderthinfalse
\newif\ifstablesinternalthin
\stablesinternalthintrue
\newif\ifstablesomit
\newif\ifstablemode
\newif\ifstablesright
\stablesrightfalse
%
%
\newdimen\stablesbaselineskip
\newdimen\stableslineskip
\newdimen\stableslineskiplimit
%
%
\newcount\stablesmode
\newcount\stableslines
\newcount\stablestemp
\stablestemp=3
\newcount\stablescount
\stablescount=0
\newcount\stableslinet
\stableslinet=0
%
%
%
\newcount\stablestyle
\stablestyle=0
%
%
\def\stablesleft{\quad\hfil}%
\def\stablesright{\hfil\quad}%
%
%
\catcode`\|=\active%
%
%
\newcount\stablestrutsize
\newbox\stablestrutbox
\setbox\stablestrutbox=\hbox{\vrule height10pt depth5pt width0pt}
\def\stablestrut{\relax\ifmmode%
                         \copy\stablestrutbox%
                       \else%
                         \unhcopy\stablestrutbox%
                       \fi}%
%
%
\newdimen\stablesborderwidth
\newdimen\stablesinternalwidth
\newdimen\stablesdummy
\newcount\stablesdummyc
\newif\ifstablesin
\stablesinfalse
%
%
\def\begintable{\stablestart%
  \stablemodetrue%
  \stablesadj%
  \halign%
  \stablesdef}%
\def\stablesadj{%
  \ifcase\stablestyle%
    \hbox to \hsize\bgroup\hss\vbox\bgroup%
  \or%
    \hbox to \hsize\bgroup\vbox\bgroup%
  \or%
    \hbox to \hsize\bgroup\hss\vbox\bgroup%
  \or%
    \hbox\bgroup\vbox\bgroup%
  \else%
    \errhelp=\stablestylehelp%
    \errmessage{Invalid style selected, using default}%
    \hbox to \hsize\bgroup\hss\vbox\bgroup%
  \fi}%
\def\stablesend{\egroup%
  \ifcase\stablestyle%
    \hss\egroup%
  \or%
    \hss\egroup%
  \or%
    \egroup%
  \or%
    \egroup%
  \else%
    \hss\egroup%
  \fi}%
\def\stablestart{%
  \ifstablesin%
    \errhelp=\stablesmultiplehelp%
    \errmessage{An S-Table cannot be placed within an S-Table!}%
  \fi
  \global\stablesintrue%
  \global\advance\stablescount by 1%
  \message{<S-Tables Generating Table \number\stablescount}%
  \begingroup%
  \stablestrutsize=\ht\stablestrutbox%
  \advance\stablestrutsize by \dp\stablestrutbox%
  \ifstablesborderthin%
    \stablesborderwidth=\stablesthinline%
  \else%
    \stablesborderwidth=\stablesthickline%
  \fi%
  \ifstablesinternalthin%
    \stablesinternalwidth=\stablesthinline%
  \else%
    \stablesinternalwidth=\stablesthickline%
  \fi%
  \tabskip=0pt%
  \stablesbaselineskip=\baselineskip%
  \stableslineskip=\lineskip%
  \stableslineskiplimit=\lineskiplimit%
  \offinterlineskip%
  \def\borderrule{\vrule width \stablesborderwidth}%
  \def\internalrule{\vrule width \stablesinternalwidth}%
  \def\thinline{\noalign{\hrule height \stablesthinline}}%
  \def\thickline{\noalign{\hrule height \stablesthickline}}%
  \def\trule{\omit\leaders\hrule height \stablesthinline\hfill}%
  \def\ttrule{\omit\leaders\hrule height \stablesthickline\hfill}%
  \def\tttrule##1{\omit\leaders\hrule height ##1\hfill}%
  \def\stablesel{&\omit\global\stablesmode=0%
    \global\advance\stableslines by 1\borderrule\hfil\cr}%
  \def\el{\stablesel&}%
  \def\elt{\stablesel\thinline&}%
  \def\eltt{\stablesel\thickline&}%
  \def\elttt##1{\stablesel\noalign{\hrule height ##1}&}%
  \def\elspec{&\omit\hfil\borderrule\cr\omit\borderrule&%
              \ifstablemode%
              \else%
                \errhelp=\stablelinehelp%
                \errmessage{Special ruling will not display properly}%
              \fi}%
  \def\stmultispan##1{\mscount=##1 \loop\ifnum\mscount>3 \stspan\repeat}%
  \def\stspan{\span\omit \advance\mscount by -1}%
  \def\multicolumn##1{\omit\multiply\stablestemp by ##1%
     \stmultispan{\stablestemp}%
     \advance\stablesmode by ##1%
     \advance\stablesmode by -1%
     \stablestemp=3}%
  \def\multirow##1{\stablesdummyc=##1\parindent=0pt\setbox0\hbox\bgroup%
    \aftergroup\emultirow\let\temp=}
  \def\emultirow{\setbox1\vbox to\stablesdummyc\stablestrutsize%
    {\hsize\wd0\vfil\box0\vfil}%
    \ht1=\ht\stablestrutbox%
    \dp1=\dp\stablestrutbox%
    \box1}%
  \def\stpar##1{\vtop\bgroup\hsize ##1%
     \baselineskip=\stablesbaselineskip%
     \lineskip=\stableslineskip%
     \lineskiplimit=\stableslineskiplimit\bgroup\aftergroup\estpar\let\temp=}%
  \def\estpar{\vskip 6pt\egroup}%
  \def\stparrow##1##2{\stablesdummy=##2%
     \setbox0=\vtop to ##1\stablestrutsize\bgroup%
     \hsize\stablesdummy%
     \baselineskip=\stablesbaselineskip%
     \lineskip=\stableslineskip%
     \lineskiplimit=\stableslineskiplimit%
     \bgroup\vfil\aftergroup\estparrow%
     \let\temp=}%
  \def\estparrow{\vfil\egroup%
     \ht0=\ht\stablestrutbox%
     \dp0=\dp\stablestrutbox%
     \wd0=\stablesdummy%
     \box0}%
  \def|{\global\advance\stablesmode by 1&&&}%
  \def\|{\global\advance\stablesmode by 1&\omit\vrule width 0pt%
         \hfil&&}%
  \def\vt{\global\advance\stablesmode by 1&\omit\vrule width \stablesthinline%
          \hfil&&}%
  \def\vtt{\global\advance\stablesmode by 1&\omit\vrule width
\stablesthickline%
          \hfil&&}%
  \def\vttt##1{\global\advance\stablesmode by 1&\omit\vrule width ##1%
          \hfil&&}%
  \def\vtr{\global\advance\stablesmode by 1&\omit\hfil\vrule width%
           \stablesthinline&&}%
  \def\vttr{\global\advance\stablesmode by 1&\omit\hfil\vrule width%
            \stablesthickline&&}%
  \def\vtttr##1{\global\advance\stablesmode by 1&\omit\hfil\vrule width ##1&&}%
  \stableslines=0%
  \stablesomitfalse}
\def\stablesdef{\bgroup\stablestrut\borderrule##\tabskip=0pt plus 1fil%
  &\stablesleft##\stablesright%
  &##\ifstablesright\hfill\fi\internalrule\ifstablesright\else\hfill\fi%
  \tabskip 0pt&&##\hfil\tabskip=0pt plus 1fil%
  &\stablesleft##\stablesright%
  &##\ifstablesright\hfill\fi\internalrule\ifstablesright\else\hfill\fi%
  \tabskip=0pt\cr%
  \ifstablesborderthin%
    \thinline%
  \else%
    \thickline%
  \fi&%
}%
\def\endtable{\advance\stableslines by 1\advance\stablesmode by 1%
   \message{- Rows: \number\stableslines, Columns:  \number\stablesmode>}%
   \stablesel%
   \ifstablesborderthin%
     \thinline%
   \else%
     \thickline%
   \fi%
   \egroup\stablesend%
\endgroup%
\global\stablesinfalse}
%
%

\chapter{Introduction}

Ever since Witten's pioneering work [\Wito,\Witt] there has been a great deal
of interest in topological field theories, both as a tool for calculating
topological invariants of manifolds and as possible physical theories
describing a "phase" of quantum gravity where there are no local gravitational
degrees of freedom. In particular, the topological sigma model [\Witt] is hoped
to describe an unbroken "phase" of string theory occurring in a high energy
limit. However, if topological theories are to be of direct physical relevance,
their topological invariance must be spontaneously broken by some, as of yet
unknown, mechanism. In order to look for symmetry breaking mechanisms, it is of
interest to  construct topological theories which contain nontrivial
potentials. In addition, the calculation of some observables in the topological
sigma model may be simplified by using a massive model in the (exact)
semi-classical limit that the mass tends to infinity. Finally, the relationship
between massive sigma models and Landau-Ginzburg theories has received
substantial interest recently; we will see here that the similarities between
these two theories are perhaps most transparent in their topological phases.

In [\Labo] an attempt was made to include potential terms for the topological
sigma model by twisting the massive (2,2) supersymmetric sigma model. However,
the potential employed there was expressed as the length of a Killing vector on
the target manifold. This form of potential cannot be defined for arbitrary
choices of the metric on a manifold with a given topology and as such is
somewhat unsatisfactory. We begin here by constructing a new
topological sigma model with potential by considering the twisted massive
(2,0) supersymmetric sigma model, for which no Killing condition on the
potential is required. We briefly discuss the discrepancies between the results
presented here and those obtained in [\Labo] and the role of central
charges in the twisted algebra. Finally we calculate the vacuum expectation
value of some observables and note that the topological massive sigma model
constructed here may be interpreted as interpolating between the topological
massless sigma model of [\Witt] and the topological Landau-Ginzburg model of
[\Ito].

\chapter{Constructing Topological Field Theories}

In general, to construct a topological field theory one needs a scalar
Grassmann
operator $Q$ such that $Q^2=0$, which we interpret as a BRST operator
representing a symmetry $\cal S$. We require that all physical states $\mid
phys >$ are invariant under $\cal S$, hence $Q\mid phys>=0$. Moreover, two
physical states are identified if they differ by a Q-exact state. Thus the
physical states correspond to Q-cohomology classes. The observables $\cal O$
are
required to satisfy $\{Q,{\cal O}\}=0$ and $\delta {\cal O} / \delta g^{\mu\nu}
= \{Q,{\cal K}_{\mu\nu}\}$ for some ${\cal K}_{\mu\nu}$, where $g_{\mu\nu}$ is
the spacetime metric and $\{ \ ,\ \}$ is a ${\bf Z}_2$ graded commutator
[\Bir].
Two observables are identified if their difference is Q-exact and therefore
they also correspond to Q-cohomology classes.

If we take the action of the theory to be a Q-commutator
$$
S=\{Q,V\}
\eqn\S
$$
for some $V$, then it follows that $\{Q,S\}=\{Q^2,V\}=0$ and the action is
invariant under $\cal S$. Furthermore, if Q does not depend on the
metric $g_{\mu\nu}$ then the energy-momentum tensor of the theory is
Q-exact
$$
T_{\mu\nu}={\delta S\over{ \delta g^{\mu\nu}}}
=\{Q,{\delta V\over{ \delta g^{\mu\nu}}}\}\ .
\eqn\T
$$
The expectation value of an observable is defined via the Feynman Path integral
as
$$
<{\cal O}> \equiv \int\! {\cal O}e^{-S} \ .
\eqn\pathint
$$
In {\pathint} we must specify what the boundary conditions
are for the various fields. While the boundary conditions for bosonic fields
are always even, those for the fermionic fields can be either
even or odd. Often however, the Grassmann
fields of a topological theory are interpreted as ghosts and so do not obey
fermionic statistics, as will be the case here. We will therefore employ
periodic boundary conditions for all the fields in {\pathint}.

It follows from the above definitions that the expectation value of an
observable is metric independent since
$$\eqalign{
{\delta  \over{\delta g^{\mu\nu}}}<{\cal O}>
&={\delta \over{\delta g^{\mu\nu}}} \int\! {\cal O}e^{-S} \cr
&=\int\! \left({\delta {\cal O}\over \delta g^{\mu\nu}}-{\cal O}T_{\mu\nu}
\right)e^{-S} \cr
&=<\left\{Q,{\cal K}_{\mu\nu} - {\cal O}{\delta
V\over{\delta g^{\mu\nu}}} \right\}> \cr
&=0 \ . \cr} \eqn\var
$$
Therefore there are no dynamics in the system. In addition, if $Q$ does not
depend on a particular coupling $g$, then the expectation values of any
observable also do not depend on $g$. This can easily be seen by an identical
argument to that used in {\var}. Assuming that the path integral measure is
topologically invariant (i.e. there are no anomalies) then if the above
conditions are satisfied we have a topological field theory.

In the
topological sigma model the target space metric $g_{ij}$ plays the role of a
coupling on which Q explicitly depends. Thus the above argument does not
suffice to prove invariance with respect to $g_{ij}$. However, if under a
metric variation V changes by $\delta V=V^{ij}\delta g_{ij}$ and $Q \mid phys>
=0$ for any choice of $g_{ij}$ used in constructing $Q$, then $\delta S$ is
always a $Q$ commutator as
$$\eqalign{
\delta S &= \{Q, \delta V\} + \{\delta Q, V\} \cr
&= \{Q, \delta V - V\} + \{Q', V\} \ , \cr}
\eqn\Qproof
$$
where $\delta Q = Q' - Q$ and $Q \mid phys> = Q' \mid phys> =0$. Hence,
provided $\delta V$ can be defined, the variation of any target space metric
invariant observable has zero expectation value.

In [\Labo] a twisted version of the massive topological
sigma model was constructed which required the manifold to admit a Killing
vector. Hence the number of manifolds for which a potential can be defined is
limited and the metric deformations must be suitably restricted. Furthermore
the
Q generator used in [\Labo] was not nilpotent. Instead $Q^2$ acts as the Lie
derivative along the Killing vector. While it is possible to define Q
cohomology when $Q^2 \ne 0$ using the methods of [\cmrlh], we will
avoid this issue here by maintaining $Q^2=0$.

\chapter{Twisting the massive (2,0) model}

The (2,0) supersymmetry algebra consists of two hermitian supersymmetry
generators $Q^a_+,\ a=1,2$ which we combine into a single complex generator
$\Q={1\over\sqrt 2}(Q^1_++iQ^2_+)$ and its hermitian conjugate
$\bQ={1\over\sqrt 2}(Q^1_+-iQ^2_+)$. We note that in two dimensions it is
possible to have left and right handed vectors, which we denote by $\ne$ and
$=$ subscripts respectively, as these are just "self-dual" and "anti-self-dual"
conditions. The subscripts, counted as $1$ and $
-1$ respectively, also indicate the "Lorentz charge"\foot{On a Riemann
surface these become holomorphic and antiholomorphic vectors respectively.}.
Together with the momentum generators $P_{\ne}$ and $P_=$ we have the (2,0)
supersymmetry algebra
$$\eqalign{
&\{\Q,\bQ \}=\Pp \ ,\ \ \ \
\ \ \Q^2 = \bQ^2 = 0 \ ,\cr &[J,\Q]={1\over2}\Q \ ,\ \ \ \ \ \ \ \
[J,\bQ]={1\over2}\bQ\ ,\cr &[J,\Pm]=-\Pm \ , \ \ \ \ \ \ \ \ \ [J,\Pp]=\Pp \
,\cr  &[U,\Q]=\Q \ ,\ \ \ \ \ \ \ \ \ \ [U,\bQ]=-\bQ \ ,\cr
&[U,J]\ =\ [U,\Pm]\ =\ [U,\Pp]\ =\ 0 \ .\cr
} \eqn\algebra
$$
Here, $J$ is the generator of two dimensional Lorentz transformations and $U$
generates an internal $SO(2)$ rotation between $Q^1_+$ and $Q^2_+$. While
only left handed generators appear in {\algebra} $U$ can be split
into left and right components $U = U_L \otimes U_R$ corresponding to the left
and right handed modes on the world sheet. Under the generators $J \otimes U_L
\otimes U_R$, $\ (\Q, \bQ)$ transform as  $({1\over2},1,0) \oplus
({1\over2},-1,0)$. If we twist {\algebra} by identifying a new generator of
Lorentz transformations
$$
T=J-{1\over2}U_L+{1\over2}U_R \ ,
\eqn\T
$$
then, with respect to $T \otimes U_L \otimes U_R$, $\ (\Q, \bQ)$ transform as
$(0,1,0) \oplus (1,-1,0)$. Thus we have a scalar generator, which we denote by
$Q$ and a left handed vector generator, which we denote by $\bQp$. The new
algebra takes the form
$$\eqalign{
&\{Q,\bQp \}=\Pp \ ,\ \ \ \ \ \ Q^2 = \bQp^2 = 0 \ ,\cr
&[T,Q]=0 \ ,\ \ \ \ \ \ \ \ \ \ \ \ \ [T,\bQp]=\bQp\ ,\cr
&[T,\Pm]=-\Pm \ , \ \ \ \ \ \ \ [T,\Pp]=\Pp \ ,\cr
&[U,Q]=Q \ ,\ \ \ \ \ \ \ \ \ \ \ \ [U,\bQp]=-\bQp \ ,\cr
&[U,T]\ =\ [U,\Pm]\ =\ [U,\Pp]\ =\ 0 \ .\cr }
\eqn\twist
$$
We have therefore identified a generator $Q$ which we may use as the BRST
operator for topological symmetry. This is the so called "A" twist. Another
possibility is the "B" twist $T = J -{1\over2}U_L-{1\over2}U_R$ in which
case the algebra {\twist} remains the same but the twisted fields
change accordingly. Before we address the problem of finding a set of fields
for the algebra {\twist} to act on, we give a short description of the massive
(2,0) supersymmetric sigma model.

The massive (2,0) supersymmetric sigma model is defined by scalar maps
$\phi^i$,
and their anticommuting spinor superpartners $\lambda_+^i$, which map from a
two
dimensional Minkowski spacetime $\Sigma$ (the base space) into an arbitrary
complex manifold $\cal M$ of (real) dimension $D$ (the target space). $\cal M$
is endowed with an hermitian metric $g_{ij}$, complex structure $J^i_{\ j}$ and
antisymmetric tensor $b_{ij}$. In addition, there is an anticommuting spinor
field $\zeta^a_-$ which maps from $\Sigma$ into an arbitrary complex vector
bundle $\ \Xi\ $ over $\cal M$ with hermitian metric $h_{ab}$ and connection
$A^a_{i\ b}$. To include a potential for the fields we introduce a section
$s^a$ of the bundle $\ \Xi\ $, which is the sum of a holomorphic and an
antiholomorphic section. The action given in complex coordinates is
$$\eqalign{
S_{(2,0)}= \int\! d^2x & \left\{
(g_{I\J}+b_{I\J})\dplus\phi^I\dminus\phi^{\J}
+ig_{I\J}\lambda^I_+\nabla^{(+)}_=\lambda^{\J}_+
-ih_{A\B}\zeta^A_-\hat{\nabla}_{\ne}\zeta^{\B}_-
\right. \cr  &\left.
-{1\over2}\zeta^A_-\zeta^{\B}_- F^{A \B}_{I \J}\lambda^I_+\lambda^{\J}_+
+mh_{A \B}\hat{\nabla}_I s^A\lambda^I_+\psi^{\B}_-
-{1\over4}m^2h_{A \B}s^As^{\B} \right\} \ .\cr }
\eqn\potential
$$
Here $F^{A \B}_{I \J}$ is the curvature of the connection $A^a_{i\ b}$,
$\hat{\nabla}$ is the covariant derivative with respect to $A^a_{i\ b}$
while $\nabla^{(+)}$ is the covariant derivative with respect to the
Levi-Civita connection with torsion $H_{ijk}={3\over2}\partial_{[i}b_{jk]}$.

In this paper we will be concerned primarily with the twisted version of the
(2,0) supersymmetric model {\potential}. We may consider the (2,2)
supersymmetric model, as was done in [\Labo], by taking the special case
[\Hulo] where we identify $\ \Xi\ $ with $T\cal M$ and $A^a_{i\ b}$ with the
spin connection $\omega_{i}^{jk}$ by introducing a vielbein $e_i^{\ a}$.
Furthermore, in the (2,2) supersymmetric case, the section must be defined by a
holomorphic Killing vector $X^i$ to be
$$
s^a = e_{i}^{\ a}(u^i-X^i) \ ,
\eqn\sa
$$
where $\partial_{[i} u_{j]}=X^k H_{ijk}$. The presence of the left
handed supersymmetries, however, induces central charges in the algebra
{\algebra}  given by the derivative of the Killing vector. Thus, with regards
to the discussion above, the (2,2) model with potential {\sa} cannot be twisted
into a topological theory as is done in [\Labo], unless $X^i=0$. In this
case (locally) $u_i=\partial_i f$ for some scalar f.
Contrary to the claim in [\Labo,\Labt], it is possible to twist this theory and
the result is a special case of the model constructed below, provided we
interpret the scalar $f$ as a worldsheet 1-form, i.e. $f \ \in \
\Phi_*(\Lambda^1(\Sigma))$. In this way Lorentz invariance is maintained in the
twisted model. In order to construct a topological version of the massive sigma
model {\potential} we will find it necessary make the identification $\Xi
\equiv T\cal M$, $A^a_{i\ b}=e_j^{\ a} e_{k\ b}\omega_{i}^{jk}$. For simplicity
we will also assume that $\cal M$ is compact in what follows. However, to
prevent the appearance of central charges and Killing vectors, we do not
enforce {\sa}. It is therefore possible to view the topological model below as
a twisted version of the massless (2,2) supersymmetric sigma model, with a
potential which breaks the supersymmetry down to (2,0).

Our next step in the construction is to replace the fields $(\phi^I,\phi^{\I},
\lambda^I_+,\lambda_+^{\I},\zeta^I_-,\zeta^{\I}_-)$ by their twisted
counterparts which we denote as
$(\phi^I,\phi^{\I},\eta^I,\psi^{\I}_{\ne},\psi^{I}_=,\eta^{\I})$. Under the
generators $T \otimes U$ these fields transform as $(0,0) \oplus (0,0) \oplus
(0,1) \oplus ({1\over2},-1) \oplus (-{1\over2},-1) \oplus (0,1)$. $\lambda^I_+$
and $\psi^{\I}_-$ are now worldsheet scalars $\eta^I$ and $\eta^{\I}$, while
$\eta^{\I}_+$ and $\zeta^I_-$ are twisted into the $\psi^{\I}_{\ne}$ and
$\psi^I_=$ components of a worldsheet 1-form, $\psi^i\  \in \
\Lambda^{1}(\Sigma) \otimes \Phi^*(T{\cal M)}$. The subscripts $=$ and $\ne$
can be viewed as referring to the (1,0) and (0,1) components of 1-forms on
$\Sigma$ respectively. Here we have performed an "A" twisting described above.
Had we used the "B" twist we would arrive at an action similar to that in
[\Labo]. The proof that the twisted version of a theory is indeed
topological requires us to explicitly write the action in the form of equation
{\S}. It is therefore necessary to find the action of Q on the twisted fields.

At this point we may generalize the construction to non Hermitian manifolds. To
this end we simply postulate a set of fields  $(\phi^i, \eta^i, \psi^i)$, where
$\phi^i, \eta^i$ are scalars and $\psi^i$ a 1-form with components $\psi^i_=$
and $\psi^i_{\ne}$, along with their transformations under Q. In order to close
the Q-algebra off-shell it is necessary to introduce a commuting, non
propagating 1-form field $H^i \in \ \Lambda^1(\Sigma) \otimes \Phi^*(T{\cal
M)}$ with components $H^i_=$ and $H^i_{\ne}$, transforming under the action of
$T \otimes U_L \otimes U_R$ as $(-1,0,0)$ and $(1,0,0)$ respectively.
Furthermore $\psi^i$ and $H^i$ are self dual in the sense that $\psi^i_= =
-iJ^i_{\ j} \psi^j_{=}$, $\psi^i_{\ne} = iJ^i_{\ j} \psi^j_{\ne}$ and similarly
for $H^i$. In the special case that $\cal M$ is an Hermitian complex manifold
these constraints are solved by setting $\psi^I_{\ne}=\psi^{\I}_==0$ whereby we
recover the above twisted fields.

The action of the generator $Q$ is the same as the standard
topological sigma model [\Witt]:
$$\eqalign{
[Q,\phi^i] &= i\eta^i\ ,\cr
\{Q,\eta^i\} &= 0\ , \cr
\{Q,\psi^i\} &=  H^{i}
+{1\over2}i\nabla_jJ^i_{\ k} \eta^j \psi^k -i\Gamma^{i}_{jk}\eta^j\psi^k\ ,\cr
[Q,H^{i}] &= {1\over2}i\nabla_k J^i_{\ j} \eta^k
H^j-i\Gamma^i_{jk}\eta^jH^k - {1\over4}(\nabla_j J^i_{\ m})(\nabla_k
J^m_{\ \ l})\eta^j\eta^k\psi^l\cr
&\ \ \ \
-{1\over2}(R^i_{\ jkl}-\nabla_k\nabla_lJ^i_{\ j})\eta^k\eta^l\psi^j \
. \cr} \eqn\Qrep
$$
The first commutator represents the symmetry of an arbitrary shift in the
coordinate $\phi^i$ of $\cal M$, while the second is necessary for $Q^2=0$. The
third anticommutator may be taken as the definition of the non propagating
field
$H^i$ while the condition $\{Q^2,\psi^i\}=0$ determines the commutator
$[Q,H^i]$ uniquely. That $[Q^2,H^i]$ vanishes follows from a lengthy but
straightforward calculation.

In the twisted algebra {\twist} we interpret the $U$ invariance of the theory
as
a ghost symmetry and call the corresponding quantum number of a field its ghost
number. Therefore, as $[Q,U]=Q$, the action of $U$ on a field raises its ghost
number by one. We wish to construct Lorentz invariant theories which preserve
the $U$ symmetry, so that both the $T$ and $J$ symmetries are preserved. Hence
we must find a topologically invariant Lagrangian with ghost number 0. In
addition, if the theory is to be topological with respect to the worldsheet
$\Sigma$, the Lagrangian must certainly be conformally invariant. The scalars
$\phi^i$ and $\eta^i$ have conformal dimension 0 while the worldsheet 1-forms
$\psi^i$ and $H^i$ have conformal dimension 1. The properties of the
various fields are summarized in table 1 below.

\vskip 1cm

\begintable
$\hbox{ Field} $ | Statistics | Conformal
Dimension | $\hbox{ Ghost Number} $ \elt
 $\phi^i $ | $+$ | $0$ | $ 0$ \elt
 $\eta^i $ | $-$ | $0$ | $ 1$ \elt
 $\psi^i$| $-$ | $1$ | $-1$ \elt
 $H^i  $ | $+$ | $1$ | $ 0$
\endtable

\centerline{{\bf Table 1:} Properties of the Twisted Fields.}

\vskip 1cm

In the (2,0) supersymmetric sigma model {\potential} the mass term breaks
conformal invariance at the classical level and so one may suspect that we can
not construct a topological version of this theory. However, in order to
maintain Lorentz invariance in the twisted
model, we must interpret $s^i$ as a tangent space valued,
worldsheet 1-form; $s^i \ \in \  \Lambda^1(\Sigma) \otimes \Phi^*(T{\cal M})$,
with components $s_{=}^i$ and $s_{\ne}^i$. In this case $s^i$ has conformal
dimension 1 and the "mass" parameter $m$ is dimensionless.

In order to construct a topological version of the theory {\potential} all we
have to do now is specify a suitable scalar function $V$ in {\S} which has
ghost number -1 and conformal dimension 0. A sufficiently general choice is
$$
V= \int\! d^2x\left\{ g_{ij}
\psi^i_= \dplus
\phi^j - \alpha g_{ij}\psi^i_= H^{j}_{\ne}
-mg_{ij}\psi^i_=s_{\ne}^j \ + (= \leftrightarrow \ne) \right\} \ ,
\eqn\Vdef
$$
where $\alpha$ and $m$ are dimensionless constants. An alternative, but
equivalent construction could have been made by redefining $H^i \rightarrow
H^i + {m\over\alpha} s^i$ in {\twist} and dropping the last term in
{\Vdef}. We must then modify the $Q$ commutators  correspondingly, but this is
easily done and automatically maintains $Q^2=0$.

The topological action derived from {\Vdef} using {\S} and {\Qrep} is
$$\eqalign{
S_{top} = \int\!
d^2x &\left\{ -2\alpha g_{ij}H_=^iH_{\ne}^j
+ g_{ij}(\dminus\phi^i-ms^i_=)H^j_{\ne}+ g_{ij}(\dplus\phi^i-ms^i_{\ne})H^j_{=}
\right. \cr
&\left.
-\alpha\psi^i_=\psi^j_{\ne}
(R_{ijkl}
- \nabla_k \nabla_l J_{ij}+{1\over2}\nabla_k J_{im} \nabla_l J^m_{\ j})
\eta^k\eta^l
\right. \cr
&\left.
-ig_{ij}\psi^i_=(\nabla_{\ne}\eta^j
+{1\over2}\nabla_kJ^j_{\ l}\dplus\phi^l\eta^k)
-ig_{ij}\psi^i_{\ne}(\nabla_{=}\eta^j
+{1\over2}\nabla_kJ^j_{\ l}\dminus\phi^l\eta^k)
\right. \cr
&\left.
-img_{ij}(\nabla_k s_{\ne}^i
+ {1\over2}\nabla_kJ^{i}_{\ m}s_{\ne}^m)\eta^k\psi^j_=
\right. \cr
&\left.
-img_{ij}(\nabla_k s_=^i
+ {1\over2}\nabla_kJ^{i}_{\ m}s_{=}^m)\eta^k\psi^j_{\ne}
\right\} \ . \cr}
\eqn\LH
$$
Here we see that $H^i$ is indeed non-propagating. If we remove $H^i$ by its
equation of motion (keeping its self duality in mind), {\LH} becomes the
more familiar sigma model action
$$\eqalign{
S_{top} = \ & \cr
\int\! d^2x &\left\{
{1\over2\alpha}(g_{ij}+iJ_{ij})\dplus\phi^i\dminus\phi^j
-\alpha\psi^i_=\psi^j_{\ne}
(R_{ijkl} - \nabla_k \nabla_l J_{ij}
+{1\over2}\nabla_k J_{im} \nabla_l J^m_{\ j})
\eta^k\eta^l
\right. \cr
&\left.
-ig_{ij}\psi^i_=(\nabla_{\ne}\eta^j
+{1\over2}\nabla_kJ^j_{\ l}\dplus\phi^l\eta^k)
-ig_{ij}\psi^i_{\ne}(\nabla_{=}\eta^j
+{1\over2}\nabla_kJ^j_{\ l}\dminus\phi^l\eta^k)
\right. \cr
&\left.
-img_{ij}(\nabla_k s_{\ne}^i
+ {1\over2}\nabla_kJ^{i}_{\ m}s_{\ne}^m)\eta^k\psi^j_=
-img_{ij}(\nabla_k s_{=}^i
+ {1\over2}\nabla_kJ^{i}_{\ m}s_{=}^m)\eta^k\psi^j_{\ne}
\right. \cr
&\left.
-{m\over2\alpha}(g_{ij}+iJ_{ij})(s_{=}^i\dplus\phi^j +s_{\ne}^i\dminus\phi^j)
+{m^2\over2\alpha}(g_{ij}+iJ_{ij})s^i_=s^j_{\ne}
\right\} \ . \cr}
\eqn\L
$$

So far we have implicitly assumed that the worldsheet metric $g_{\mu\nu}$ is
flat. However, as all the formulas we have written are in terms of differential
forms and the action is conformally invariant, we can extend the model to be
defined on an arbitrary Riemann base manifold $\Sigma$ [\Witt]. Furthermore, as
is the case with the massless topological sigma model [\Witt], it is not
necessary for the Nijenhuis tensor to vanish in order to close the topological
algebra {\Qrep}. Hence the manifold $\cal M$ need only be almost complex. In
addition, no restrictions are required on the section $s^i$. We may therefore
interpret {\L} as a more general model defined for an almost complex manifold
$\cal M$. This generalized model cannot arise from twisting the massive
sigma model {\potential}, since the supersymmetry algebras {\algebra} and
{\twist} are no longer satisfied.

Consider the case where $\cal M$ is complex with an Hermitian metric. The
action
{\L} then reduces to a twisted version of the model {\potential} (with
$b_{ij}=0$ and $\Xi$ identified with $T^*{\cal M}$), with an additional mass
term $ {m\over2\alpha}g_{I \J}(s_{=}^I\dplus\phi^{\J} +
\dminus\phi^Is_{\ne}^{\J})$. The mass terms therefore do not entirely arise
from simply twisting the massive sigma model {\potential}. We will see below
however, that the appearance of this additional mass term allows the
topological massive sigma model to be identified with the topological
Landau-Ginzburg model in the limit $m \rightarrow \infty$. The Q commutators
{\Qrep} rely on the identification of $\ \Xi\ $ with $T^*{\cal M}$ so that the
massless version of {\potential} admits (2,2) supersymmetry. We may arrive at a
model related to the twisted (2,0) supersymmetric sigma model by setting
$\psi^I_= = \eta^{\I} = 0$. We shall discuss this case in more detail in the
next section. In all these cases by
setting $m=0$ we obtain topological twisted versions of
the (2,2) and (2,0) supersymmetric sigma sigma models first constructed in
[\Witt]. In the special case that $s^i=\partial^if$ where $f$ is a 1-form on
$\Sigma$ pushed forward to $\cal M$ by $\phi^i$, the action {\L} is related to
a twisted version of the (2,2) supersymmetric sigma model {\potential} with the
Killing vector equal to zero.

It is clear from the discussion earlier that the stress-energy
tensor of this theory is Q-exact and is given by $T_{\mu\nu}=\{Q,\delta V /
\delta g^{\mu\nu}\}$. Hence the theory is topological with respect to the
worldsheet metric. For completeness we also show explicitly that $\delta S
/\delta g_{ij}$ is Q-exact so that the theory is topological with respect to
the target space metric. A straight forward calculation shows that
$$
\delta S = \left\{Q, \int\! d^2x (\psi^i_=\partial_{\ne}\phi^j
-m\psi^i_=s_{\ne}^j \delta g_{ij} + = \leftrightarrow \ne\ ) \right\} \ .
\eqn\Svar
$$

\chapter{Observables}

In the massless topological sigma model an interesting class of observables can
be defined
by an n-form $A_{i_1...i_n}d\phi^{i_1}...d\phi^{i_n}$ on $\cal M$ [\Witt] by
$$
{\cal O}_{A}^{(0)} = A_{i_1...i_n}\eta^{i_1}...\eta^{i_n}\  .
\eqn\Omassless
$$
Then we have $\{Q,{\cal O}_A^{(0)}\}=-i{\cal O}_{d_{\cal M}A}^{(0)}$, where
$d_{\cal M}$ is the exterior derivative on $\cal M$. Hence ${\cal O}_{A}^{(0)}$
is a  BRST observable if and only if $d_{\cal M}A=0$. Furthermore, if two
n-forms lie in the same cohomology class, they represent the same observable.
Therefore the observables  ${\cal O}_{A}^{(0)}$ are in a one to one
correspondence with the cohomology group $H^n({\cal M})$.

In addition the (closed) n-form $A$ defines two more observables  ${\cal
O}_{A}^{(1)}$ and ${\cal O}_{A}^{(2)}$ defined over 1-cycles and 2-cycles in
$\Sigma$ respectively.  They are defined as
$$\eqalign{
{\cal O}_{A}^{(1)}& = nA_{i_1...i_n}d\phi^{i_1}\eta^{i_2}...\eta^{i_n}	 \ , \cr
{\cal O}_{A}^{(2)}& =
{n(n-1)\over2}A_{i_1...i_n}d\phi^{i_1}d\phi^{i_2}\eta^{i_3}...\eta^{i_n}	 \ ,
\cr}
\eqn\otherO
$$
It is not hard to show [\Witt] that $d_{\Sigma}{\cal O}_{A}^{(k)}$ is Q-exact,
where $d_{\Sigma}$ is the exterior derivative on $\Sigma$, and $\{Q,{\cal
O}_A^{(k)}\}=-i{\cal O}_{d_{\cal M}A}^{(k)}=0,\  k=1,2$. Therefore, by
integrating over a 1-cycle $\gamma$ and 2-cycle $\beta$ in $\Sigma$, one
obtains the observables
$$\eqalign{
{\cal W}_1(\gamma) &= \int_{\gamma}	{\cal
O}_{A}^{(1)}  \  ,  \cr {\cal W}_2(\beta) &= \int_{\beta} {\cal O}_{A}^{(2)}
\  .  \cr} \eqn\Wdef
$$
It also follows that for $k=1,2$
$$
{\cal W}_k(\partial\gamma) = \int_{\partial\gamma}	{\cal O}_{A}^{(k)}
= \int_{\gamma}	d_{\Sigma}{\cal O}_{A}^{(k)}
\eqn\proof
$$
is Q-exact. Hence the
expectations values of ${\cal W}_1(\gamma)$ and ${\cal W}_2(\beta)$
depend only on the de Rahm cohomology class of $A$ and the homology classes of
the cycles $\gamma$ and $\beta$ respectively.

It follows from the arguments in section 2 that the expectation values of the
observables are independent of the parameters $\alpha$ and $m$. Therefore we
can take the limit $m \rightarrow 0$ and recover massless topological sigma
model, or alternatively, take $m \rightarrow \infty$, where the action
simplifies and observables can be more readily calculated.

To this end we will now compute the expectation values of the observables
${\cal O}^{(0)}_A$ explicitly using the mass term to simplify the work. First
let us consider the "(2,0) model" in complex coordinates where $g_{I \J}$ is
Hermitian and we set
$\psi^I_= = H^I_= = \eta^{\I}=0$. For further simplicity we work in
the $\alpha=0$ "gauge":
$$\eqalign{ S_{top} = \int\! d^2x &\left\{
g_{I \J}(\dminus\phi^I-ms^I_=)H^{\J}_{\ne}
-ig_{\I J}\psi^{\I}_{\ne}\nabla_{=}\eta^J  \right. \cr
&\left.
-img_{I \J}\nabla_{K} s_{=}^I\eta^K\psi^{\J}_{\ne}
\right\} \ . \cr}
\eqn\LA
$$
The expectation value of ${\cal O}^{(0)}_A$ is defined as
$$
<{\cal O}^{(0)}_A> =
\int\! d[H]d[\phi]d[\eta]d[\phi]{\cal O}^{(0)}_Ae^{-S_{top}}\ .
\eqn\Ovev
$$

To evaluate {\Ovev} we may invoke the use of a Nicoli map
[\Bir], by making a change of variables to
$$
\pi^I_= = \partial_=\phi^I - ms_=^I
\eqn\Nic
$$
We must be careful here not to include any zero modes
to insure that the transformation is well defined. Therefore we leave
out the zero modes $\phi^I_0$ and integrate over them separately. We
shall elaborate on them shortly. The above transformation has the effect of
trivializing the first term in {\LA} to $g_{I \J}\pi^I_=H^{\J}_{\ne}$ and
introducing a Jacobian factor $\mid {\rm det'} (B^I_{=
\ J}) \mid $, where
$B^I_{=\ J}=\delta \pi^I_= / \delta \phi^J$, into the measure:
$$
<{\cal O}^{(0)}_A>=\int_{M} d\phi_0\int\! d[H]d[\psi]d[\eta]d[\pi]
{{\cal O}^{(0)}_A \over \mid {\rm det'} (B^I_{=\ J}) \mid}
e^{-S_{top}}\ .  \eqn\det
$$
In {\det} $M$ is the moduli space of bosonic zero modes of $B^I_{=\ J}$ and the
prime indicates that we omit the zero modes in calculating the determinant. The
Jacobian can be found as the first order term of $\pi^I_=$ in a background
field expansion of $\phi^I$ [\Bir]. This yields
$$\eqalign{
\phi^I &\rightarrow \phi^I + \xi^I \cr
\pi^I_= &\rightarrow \partial_=\phi^I - ms_=^I +
\nabla_{=}\xi^I - m \nabla_J s_{=}^I\xi^J + O(\xi^2)\ , \cr}
\eqn\expand
$$
hence
$$
B^I_{=\ J} = {\delta \pi^I_=\over \delta \xi^J} =
\nabla_{=}\delta^I_{\ J} - m \nabla_Js_{=}^I \ .
\eqn\B
$$

Performing the $H_{\ne}^I$ integration, we obtain a delta function which
projects the $\pi^I_=$ integration down on to the space of instanton solutions
$$
\partial_= \phi^I - m s_=^I = 0 \ .
\eqn\moduli
$$
Furthermore, if we now use the freedom to take the limit $m \rightarrow
\infty$, the $\phi^I$ fields become localized at the zeros of $s_=^I$. If we
assume $s_=^I$ has discrete zeros, then the integral over $\pi^I_=$ becomes a
sum over the zeros of $s_=^I$. We therefore have
$$\eqalign{
&<{\cal O}^{(0)}_A> \cr
&=
\sum_{zeros} \int_{M} d\phi_0 \int\! d[\psi]d[\eta]{{\cal O}^{(0)}_A\over \mid
{\rm det'} (B^I_{=\ J})\mid}
exp(i\int\!d^2x g_{\I J}\psi^{\I}_{\ne}(\nabla_{=}\delta^{J}_{\ K}-
m \nabla_Ks_{=}^{J})\eta^K ) \ . \cr}
\eqn\w
$$

Before continuing with the calculation we should make some remarks about zero
modes and the ghost number anomaly. As we have just seen there are potentially
bosonic zero modes $\phi^I_0$ of $B^I_{=\ J}$. In addition there may also be
fermionic zero modes $\eta^I_0$ and $\psi^I_{\ne 0}$. Indeed, the integrand in
{\w} is just $g_{\I J}\psi^{\I}_{\ne}B^{J}_{=\ K}\eta^K$. Now $B^J_{=\ K}$
and its adjoint define maps
$$\eqalign{
B&:\Phi^*T{\cal M} \rightarrow
\Lambda^{(1,0)}(\Sigma) \otimes \Phi^*T{\cal M} \ , \cr
B^{\dag} &: \Lambda^{(0,1)}(\Sigma) \otimes \Phi^*T{\cal M}
\rightarrow \Phi^*T{\cal M} \ , \cr}
\eqn\diagram
$$
which will generally have $\eta^I_0$ and $\psi^{\I}_{\ne 0}$ zero modes
respectively.  The number of these modes will in general depend upon the
1-form $s$ and the topology of $\Sigma$. In order not to commit ourselves
to a particular model, we will not discuss in any more
detail here the existence of infinitesimal fermionic zero modes.

There is, however, a global obstruction to constructing finite zero modes from
the infinitesimal ones above [\Bir]. This causes the number of finite $B^I_{=\
J}$ zero modes to be given by the index
$$
{\rm ind}B ={\rm dim Ker}B - {\rm dim
Ker}B^{\dag} \ ,
 \eqn\index
$$
which is interpreted as the virtual dimension of the moduli space $M$ and is
not, in general, a constant over $M$. The effect of the ghost number anomaly
{\index} is to give non vanishing expectation values to observables with non
zero ghost number. As the observable {\Omassless} has ghost number $n$, we need
$n$ $\eta^I_0$ modes in the path integral to obtain a non vanishing vacuum
expectation value. In addition, for these observables the presence of any
$\psi^{\I}_{\ne 0}$ modes would cause the integral {\Ovev} to vanish. Hence we
will assume that there are only $\phi^I$ and $\eta^I$ zero modes.

Continuing with the calculation and integrating over the anticommuting non zero
modes in {\w} we obtain, by a standard result of Grassmann integration,
$$\eqalign{
<{\cal O}^{(0)}_A>&=\sum_{zeros} \int_{\hat {M}} d\eta_0\ d\phi_0
{{\rm det'}(B^I_{=\ J})\over\mid {\rm det'} (B^I_{=\ J})\mid }{\cal O}^{(0)}_A
\cr &
=\sum_{zeros}\int_{\hat {M}} d\eta_0\ d\phi_0\  {\rm sgndet'}(B^I_{=\ J})
A_{I_1...I_n}(\phi_0)\eta^{I_1}_0...\eta^{I_n}_0 \ , \cr}
\eqn\al
$$
where $\hat M$ is the supermoduli space which includes the anticommuting fields
and can be viewed as the tangent bundle to $M$ since we are assuming there are
no $\psi^{\I}_{\ne}$ zero modes. As $m \rightarrow \infty$, $B^I_{=\ J}$ is
dominated by the mass term. If we deform $g_{I \J}$ so that it is flat near
the zeros of $s_=^I$ we obtain
$$\eqalign{
<{\cal O}^{(0)}_A>&=\sum_{zeros}{\rm sgndet'}(\partial_Js_{=}^I)\int_{\hat
{M}} d\eta_0\ d\phi_0 A_{I_1...I_n}(\phi_0)\eta^{I_1}_0...\eta^{I_n}_0\cr
&=\sum_{zeros}{\rm sgndet'}(\partial_Js_{=}^I) \int_{M_n}
d\phi_0 \ A_{I_1...I_n}(\phi_0) \ , \cr }
\eqn\Obs
$$
where $M_n$ is the n dimensional component of the moduli space $M$ and we have
used the canonical measure $d\eta_0=d\eta^{I_1}_0...d\eta^{I_n}_0$ for the
Grassmann integral. What are we to  make of the expression {\Obs}? In the limit
that $m \rightarrow \infty$, the zero modes of $B^I_{\ J}$ are generated by the
directions in $\cal M$, at a given zero, along which $s^I_{=}$ is flat. We
therefore interpret the integral in {\Obs} to be over the n dimensional
submanifold of $\cal M$ generated by the n flat directions of $s^I_{=}$ at each
zero.

 From {\Obs} one can read off the partition function $Z$ by considering the n=0
case, with  $A(\phi)\equiv 1$ and no zero modes. Then in
{\Obs}, the left hand side is just $Z$, while the right hand side simply
becomes a weighted sum over the zeros of $s^I_=$
$$\eqalign{
Z&=\sum_{zeros} {\rm sgndet}(\partial_J s_=^I) \cr
&=\chi({\cal M}) \ ,}
\eqn\Z
$$
which, by the Hopf Index Theorem, is just the Euler number of $\cal M$.

Having performed the above calculation it is effortless to consider the (2,2)
case where $\psi^I_=, H^I_=, \eta^{\I} \ne 0$. Here we simply repeat the
calculation with $(\psi^{\I}_{\ne},H^{\I}_{\ne},\eta^I) \leftrightarrow
(\psi^{I}_{=},H^{I}_{=},\eta^{\I})$ and combine the two results. Equation
{\Obs}
becomes
$$
<{\cal O}^{(0)}_A> =\sum_{zeros}{\rm sgndet'}(\partial_js^i) \int_{M_n}
d\phi_0 \ A_{i_1...i_n}(\phi_0) \ ,
\eqn\Obstwo
$$
where ${\rm sgndet'}(\partial_js^i) = {\rm
sgndet'}(\partial_{\J}s_{\ne}^{\I}) {\rm sgndet'}(\partial_Js_{=}^I)$ and we
sum over the zeros of $s^i$ (i.e. the common zeros of $s^I_=$ and
$s^{\I}_{\ne}$).

Thus we have reduced the calculation of $<{\cal O}^{(0)}_A>$ to {\Obs} and
recovered the standard result that the partition function is equal to  the
Euler number of $\cal M$. In the case where $s^I =
\partial^I f$ for some $f$, so that the sigma model
{\potential} possesses (2,2) supersymmetry, our assumptions about discrete
zeros
and no zero modes are then just that $f$ is a Morse function on $\cal M$ (i.e.
it
has discrete, nondegenerate extrema). The result {\Z} is then related to the
Morse
formula for the Euler number, {\it viz}
$$\eqalign{
Z= \chi({\cal M})&=\sum_{extrema}{\rm sgndet}(\partial_J\partial^If) \cr
&=\sum_{det >0} 1  \ -\  \sum_{det <0} 1 \cr
&=\sum_{n=0}^{D/2} M_{2n} - \sum_{n=0}^{D/2} M_{2n+1} \cr
&=\sum_{n=0}^{D} (-1)^n M_n \ , \cr}
\eqn\Morse
$$
where $M_n$ is the nth Morse number (i.e. the number of extrema of $f$ with n
negative modes) and $D$ is the (real) dimension of $\cal M$.

Finally, it is instructive to rewrite the action {\L}, when $\cal M$
is complex and $g_{I \J}$ Hermitian, as
$$\eqalign{
S_{top} = \int\! d^2x &\left\{
{1\over\alpha}g_{I \J}(\dminus\phi^{I} - ms_=^{I})
(\dplus\phi^{\J}-ms_{\ne}^{\J})
\right. \cr &\left.
-ig_{\I J}\psi^{\I}_{\ne}(\nabla_{=}\eta^{J} - m\nabla_{K}s^{J}_{=}\eta^{K})
-ig_{I \J}\psi^{I}_=(\nabla_{\ne}\eta^{\J} - m\nabla_{\K}s^{\J}_{\ne}\eta^{\K})
\right. \cr &\left.
+\alpha\psi^I_=\psi^{\I}_{\ne}R_{I \I J \J} \eta^J\eta^{\J}
\right\} \ . \cr}
\eqn\Lnew
$$
 From this it becomes clear that the path integral is dominated, for small
$\alpha$, by the instanton solutions {\moduli}. If we ignore the connection
terms in {\Lnew} and identify $(\phi^I,\phi^{\I},\eta^I, \eta^{\I},
\psi^I_=,\psi^{\I}_{\ne})$ with the fields $(U^I, U^{\I}, \chi^I,\chi^{\I}
\rho^I_z,\rho^{\I}_{\bar z})$ of reference [\Ito] we arrive at
the (2,2) supersymmetric topological Landau-Ginzburg model of reference [\Ito],
with the potential $W$ given by $f$ and suitably interpreted as a world sheet
1-form. In the case where we set $\psi^I_==\eta^{\I}=0$, the model is a twisted
form of the (2,0) Landau-Ginzburg model. Thus the topological Landau-Ginzburg
theories arise simply from the topological massive sigma model in the limit
where
the target space can be considered flat. In fact if the $\cal M$ admits a flat
metric (i.e. if it has the topology of ${\bf C}^{D/2}$, possibly with points
removed) then we may make the "gauge" choice $g_{ij}=\delta_{ij}$ in which case
the massive topological sigma model is simply the topological Landau-Ginzburg
theory.

In [\Ito,\With] it was shown that the observables of the topological
Landau-Ginzburg model and topological massless sigma model both have the same
representation in bosonic fields as the primary fields of N=2 superconformal
field theories. Above we calculated the expectation values of some observables
of the topological sigma model using the freedom to let $m \rightarrow \infty$.
This had the effect of allowing us to ignore the metric structure of the target
manifold (although the fields were still constrained to lie in the target
space). However, when we ignore the target space metric, the model becomes the
topological Landau-Ginzburg theory. Thus the topological massive sigma model
constructed here can be viewed as interpolating between the massless
topological sigma model of [\Witt] at $m=0$ and the topological Landau-Ginzburg
model of [\Ito] as $m \rightarrow \infty$. In the infinite mass limit the
$\phi^i$ are localized about the zeros of $s^i$ which are the vacuum states of
the corresponding Landau-Ginzburg model. Furthermore, in order that the
observables ${\cal O}_{A}$ have a non zero expectation value, $s^i$ must
possess "flat" directions. In the Landau-Ginzburg model these flat
directions produce the massless excitations which are needed for the
associated conformal field theory to be nontrivial.

\chapter{\bf Comments}

In this paper we constructed a topological massive sigma model related to the
twisted massive (2,0) supersymmetric sigma model. In addition to providing non
trivial potentials for the bosonic fields, the inclusion of mass terms in the
topological sigma model allows one to simplify the calculation of some
observables. Furthermore, we argued that in the limit $m \rightarrow \infty$
the topological massive sigma model becomes a topological Landau-Ginzburg
model. Hence the topological sigma model and the topological Landau-Ginzburg
model may be viewed as the same theory, interpolated by the topological massive
sigma model. We have only presented one possible way of constructing
topological massive sigma models here. It would be of interest to pursue other
massive models, in particular (2,0) models where the vector bundle is not
associated with the tangent space. In addition it would be interesting to
couple the model here to 2-D topological gravity and investigate the
resulting string theory and "space of all 2D topological field
theories".

The author would like to thank G. Papadopoulos and P.K. Townsend for helpful
comments and Trinity College Cambridge for financial support.

\refout

\end